\pdfoutput=1 

\documentclass{appolb}

\usepackage{amsmath,amssymb,bbm,color}
\usepackage{amsbsy}
\usepackage{euscript}
\usepackage{graphicx}
\usepackage{hyperref}
\usepackage{cite}
\usepackage{enumerate}

\newcommand{\Reu}{\EuScript{R}}

\newcommand{\Meu}{\EuScript{M}}

\newcommand{\Kbbm}{\mathbbm{K}}

\newcommand{\veps}{\varepsilon}
\newcommand{\alfb}{\bar{\alpha}}
\newcommand{\betb}{\bar{\beta}}

\newcommand{\msbar}{\overline{MS}}
\newcommand{\krk}{{\tt KRK}}

\newcommand{\krknlo}{{\textsf{KrkNLO}}}

\newcommand{\mcatnlo}[1]{\textsf{MC@NLO#1}}

\newcommand{\powheg}[1]{\textsf{POWHEG#1}}

\newcommand{\lo}{\text{LO}}
\newcommand{\nlo}{\text{NLO}}

\newcommand{\qbar}{{\bar q}}

\newcommand{\fr}{{\leftarrow}}
\newcommand{\qb}{{\bar{q}}}

\newcommand{\II}{{$\EuScript{II}$}}
\newcommand{\IF}{{$\EuScript{IF}$}}
\newcommand{\FI}{{$\EuScript{FI}$}}
\newcommand{\FF}{{$\EuScript{FF}$}}

\newcommand{\F}{{_{F}}}
\newcommand{\B}{{_{B}}}

\newcommand{\order}[1]{${\cal O}(#1)$}

\begin{document}
\eqsec  
\title{%
\vspace{-4.0cm}
\begin{flushright}
		{\small {\bf IFJPAN-IV-2020-01}}   \\
\end{flushright}
\vspace{0.5cm}
\bf On the universality of the KRK factorization scheme%
\thanks{Presented at XXV Cracow EPIPHANY Conference, January 7-10, 2020\\
This work is partly supported by
 the Polish National Science Center grant 2016/23/B/ST2/03927.
}%
}
\author{Stanis\l{}aw Jadach
\address{Institute of Nuclear Physics, Polish Academy of Sciences,\\
         31-342 Krak\'ow, ul. Radzikowskiego 152, Poland}
}
\maketitle
\begin{abstract}
The factorisation scheme (FS) abbreviated as KRK FS
including a new definition of the PDFs for initial hadrons
was formulated while developing KrkNLO scheme of matching
QCD NLO corrections for the hard process with the parton shower
heavy boson production in hadron-hadron collision and for deep inelastic
lepton-hadron scattering.
KRK FS (originally called Monte Carlo FS) can be regarded as a variant 
of the $\overline{MS}$ system.
It is therefore trivially universal, that is process independent.
The question of its universality is formulated differently:
As the basic role of KRK FS is to simplify drastically NLO corrections,
the question is now whether the same {\em single} variant of PDFs
in the KRK FS is able to achieve the same maximal simplification
of the NLO corrections for {\em all processes} with one or two initial hadrons
and any number of the final hadrons?
Our answer is positive and the proof is elaborated in the present note within
the Catani-Seymour subtraction methodology.
KRK FS is mandatory in the KrkNLO  method of matching 
NLO calculation and parton shower
-- a much simpler alternative of POWHEG and/or MC\@ NLO.
However, the use of KRK FS and the corresponding PDFs
simplifies NLO calculations for 
any other method of calculating NLO corrections and for arbitrary processes as well.
\end{abstract}
\PACS{12.20.-m, 1470.Fm}


\section{Introduction}

The first idea of the KRK factorization scheme (KRK FS)
of the \krknlo{} method of upgrading hard process
of the parton shower Monte Carlo (MC) to NLO level
was formulated for the Drell-Yan (DY) process in Ref.~\cite{Skrzypek:2011zw}.
Later on,  in Ref.~\cite{Jadach:2011cr},
the \krknlo{} method was elaborated in a quite detail
for the DY and  the deep inelastic $ep$ scattering (DIS) processes
with parton distribution functions (PDFs) defined in the KRK FS.
The first practical implementation of \krknlo{} methodology
for the DY process on top of SHERPA and HERWIG parton shower MCs was
presented in Ref.~\cite{Jadach:2015mza}, including
comparisons with the NLO and NNLO fixed order calculations,
and also comparing with the calculation in the MC@NLO~\cite{Frixione:2002ik}
and POWHEG~\cite{Nason:2004rx} matching schemes.

Later on, in Refs.~\cite{Jadach:2016acv} the use of
PDFs in the KRK factorization scheme
was formulated for the DY and Higgs production processes
and finally applied for the MC simulations of the Higgs boson production 
at the LHC within the KrkNLO method in Ref.~\cite{Jadach:2016qti}.

Universality of PDFs (process independence) is of paramount practical
importance, because it allows to determine them in one process 
(typically DIS) and then use them as an input
in order to obtain precise theoretical predictions
in any other process, with one or two incoming hadrons.
PDFs in the $\msbar$ scheme are universal, 
as we know both from experimental tests
and also from theoretical arguments.

In most the above mentioned works PDFs in the KRK FS were defined
in the context of the DY-like processes like $Z$ boson or Higgs boson
production in the $pp$ colliders, sometimes also for the DIS process.
Hence the question of the universality (process independence)
of PDFs in the KRK FS was not a burning issue but was waiting for answer.
In the present note we are going argue that one can answer this
question in a systematic way within the framework
of the Catani-Seymour subtraction scheme~\cite{Catani:1996vz}
of NLO calculations for any scattering process 
with any number of leptons and coloured
partons in the initial and final state.

Master formula for NLO calculation for $m$ partons
within the Catani-Seymour (CS) scheme~\cite{Catani:1996vz}
reads schematically as follows:
\begin{equation}
\begin{split}
&\sigma^{NLO}(p) = \sigma^{B}(p)+
\\&
 +\int_m \big[ d\sigma^{V}(p) +d\sigma^{B}(p) \otimes\; {\bf I} \big]_{\veps=0}
 +\int dz \int_m \big[ d\sigma^{B}(zp) \otimes\; ( {\bf P+K })(z) \big]_{\veps=0}
\\&
   +\int\limits_{m+1} \Big[ 
    d\sigma^{R}(p)_{\veps=0} 
   -\big(\sum_{dipoles} d\sigma^{B}(p) \otimes\; dV_{dipole}\big)_{\veps=0}
   \Big],
\end{split}
\label{eq:CSmaster}
\end{equation}
where $p$ stands for an initial parton(s) embedded in PDF(s),
symbol $\otimes$ denotes phase space convolution, colour and spin summations.
The counterterm $ d\sigma^{B}(p) \otimes\; dV_{dipole} $ defined in 
$m+1$-particle phase space
encapsulates all soft and collinear singularities --
it is added and subtracted.
Thanks to clever kinematic mapping it factorizes off
and is integrable analytically in $d=4+2\veps$ dimensions,
${\bf I} = \sum\limits_{dipoles} \int_1 dV_{dipole} $
over the entire NLO phase space.

In refs.\cite{Jadach:2011cr,Jadach:2016acv} it was shown that thanks
to transformation of PDFs from $\msbar$ to MC FS
one can get rid of the annoying third term in eq.~\ref{eq:CSmaster}
with $ ({\bf P+K }) $ matrix
for the DY-type process and DIS process.
The eliminated term collects technical artifacts of the dimensional regularization
(collinear remnants), which can be regarded as {\em unphysical}.
The resulting NLO formula reads as follows:
\begin{equation}
\begin{split}
\sigma^{NLO}(p) &= \sigma^{B}(p)
 +\int_m \big[ d\sigma^{V}(p) +d\sigma^{B}(p)\; I(\veps) \big]_{\veps=0}
\\&
   +\int\limits_{m+1} \Big[ 
    d\sigma^{R}(p)_{\veps=0} 
   -\big(\sum_{dipoles} d\sigma^{B}(p) \otimes\; dV_{dipole}\big)_{\veps=0}
   \Big].
\end{split}
\label{eq:KRKmaster}
\end{equation}
The \krknlo\ method of matching NLO calculation with PS MC
relies vitally on the validity of the above simplified formula.

\begin{figure}
\centering
{\includegraphics[width=90mm]{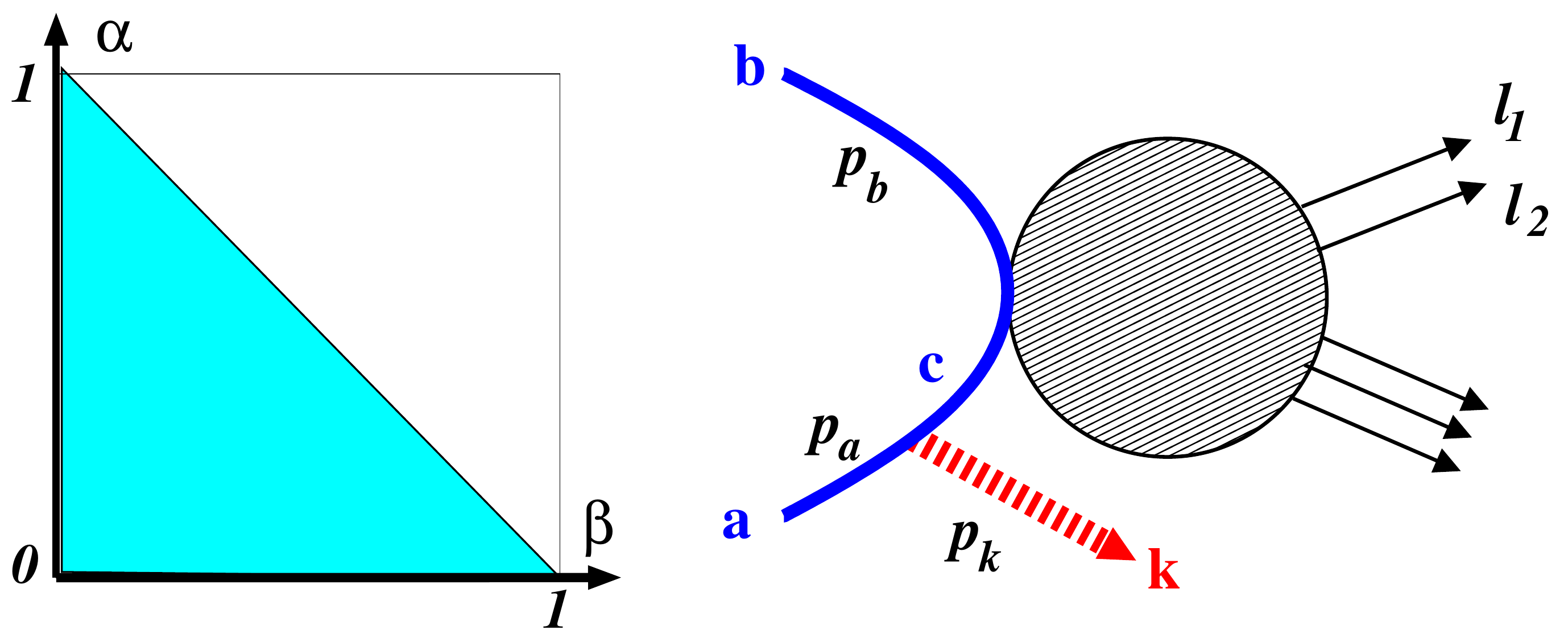}}
\caption{\sf Kinematics and Sudakov plane for DY-like processes.}
\label{fig:DYsudak}
\end{figure}

The question addressed in the following will be at the two levels:
Is the above simplification restricted to
processes with only two coloured legs, 
like heavy boson(s) production in $pp$ collision or $ep$ scattering?
Or it can be achieved for any process with arbitrary number of
coloured legs?
In case the simplification is feasible 
for any process, then the second question is:
is this {\em the same} set of PDFs in new KRK FS,
which provides for the simplification of Eq.~\ref{eq:KRKmaster}
for any process, without the need of
adjusting the definition of PDFs in the KRK FS process by process?
Full universality of the PDFs in the KRK FS requires positive
answer to both above questions.

Let us illustrate the main points of the proposed factorization
scheme and explain its role in the \krknlo\ method using examples
of the production of any heavy boson like $Z,\gamma,W,H$  in quark-antiquark annihilation
with kinematics depicted in Fig.~\ref{fig:DYsudak}.
For the sake of simplicity let us focus on the gluonstrahlung subprocesses,
i.e. $a=q,\; b=\bar{q},\; k=G, c=q$ in Fig.~\ref{fig:DYsudak}.
We are going to show why KRK FS is mandatory for \krknlo\ scheme
and what is the relation between CS dipoles and transformation
between PDFs in KRK and $\msbar$ schemes.

In the \krknlo\ matching the NLO corrected differential cross section in
the CS subtraction scheme is compared (matched) with the same distribution in 
the parton the shower with NLO corrected hard process.
Identifying and matching the same elements in both distributions
can only be successful if both of them are brought to the same form.
Following closely Ref.~\cite{Jadach:2015mza} let us compare both distributions
in the formulation without any resummation (always present in the parton shower)
and with subtraction like it is in the final CS formula in $d=4$ dimensions.

The final formula for the NLO cross section with CS dipole subtractions 
in $d=4$ dimensions reads in the notation of Ref.~\cite{Jadach:2015mza}%
\footnote{See formula of eq.~(B.7) in the notation introduced
 in eqs.~(3.1-3.7)  in  Ref.~\cite{Jadach:2015mza}.}
as follows:
\begin{equation}
\begin{split}
&\sigma_{\nlo}^{\msbar}= 
\int d x_\F d x_\B dz\; dx\; \delta_{x=z x_\F x_\B}
\Big\{
 \Big[
 \delta_{1=z} (1+\Delta_{VS})\;
\\&
+2\frac{\alpha_s}{2\pi}  P_{qq}(z) \ln\frac{\hat{s}}{\mu_F^2} 
+\Sigma_{q}(z)
\Big]
 d\sigma_0(szx,\hat\theta)\; J_{LO}
\\&
+ \Big(  d^5\sigma_1^\nlo\; (sx,\alpha,\beta, \Omega) \;J_{NLO}
       -\big( d^5\sigma_1^\F +d^5\sigma_1^\B 
  \big) 
  J_{LO}
  \Big)
\delta_{1-z=\alpha+\beta}
\Big\}
\\&\times
f^{\msbar}_q(sx, x_{\F}) f^{\msbar}_{\qb}(sx, x_{\B}).
\end{split}
\label{eq:MsbarMaster}
\end{equation}
where $J_{NLO}\equiv  J(x_\F, x_\B,z,k^T_1)$ 
and   $J_{LO}\equiv   J(x_\F, x_\B,1,0)$ are explicit
experimental event selection functions.
Two CS dipoles with initial state emitter and initial state spectator are%
\footnote{These are $d=4$ versions. 
 It is essential to define CS dipoles in $d=4+2\veps$ as well.
}
\begin{equation}
d^5\sigma^\F_{q\qbar}=d^5\sigma^{\lo}_{q\qbar}\; \frac{\alpha}{\alpha+\beta},\quad
d^5\sigma^\B_{q\qbar}=d^5\sigma^{\lo}_{q\qbar}\; \frac{\beta}{\alpha+\beta},
\label{eq:dsig5LO}
\end{equation}
where
\begin{equation}
d^5\sigma^{\lo}_{q\qbar}(sx,\alpha,\beta, \Omega)
=\frac{C_F \alpha_s}{\pi}\;
 \frac{d\alpha d\beta}{\alpha\beta}\;
 \frac{d\varphi}{2\pi} d\Omega\;
 \frac{1+(1-\alpha-\beta)^2}{2}
 \frac{d\sigma_{0}}{d\Omega}\big(sx,\hat\theta \big).
\end{equation}
Finally,  the NLO 1-real gluon emission distribution $d^5\sigma_1^\nlo $ 
is that of eq.~(3.3) in Ref.~\cite{Jadach:2015mza}
and $\Sigma_{q}(z)$, see eq.~(B.5) therein, reads
\begin{equation}
2\Sigma_q(z)= \frac{2C_F \alpha_s}{\pi}
 \bigg\{
    \frac{1+z^2}{2(1-z)} \ln\frac{(1-z)^2}{z}
   +\frac{1+z^2}{2(1-z)} \ln\frac{\hat{s}}{\mu^2}
   +\frac{1-z}{2}
 \bigg\}_+.
\end{equation}

In the \krknlo\ method upgrade of the hard process to NLO level
is done by means reweighting each MC event of the parton shower (PS)
with the single finite positive correcting weight
\[
   W^{(1)}_{NLO}(k_1),
\]
where $k_1$ is momentum of gluon with the highest transverse momentum $k_T$,
even if the PS is actually not based on the $k_T$ ordering algorithm.
The actual form of $W^{(1)}_{NLO}(k_1)$ will result from the matching procedure.
Bringing NLO corrected parton shower distribution to exactly the same
analytical formula as in eq.~(\ref{eq:dsig5LO}) is a quite nontrivial task.
It was done quite carefully and explicitly in Section 3.4 
in Ref.~\cite{Jadach:2015mza}.
The resulting formula, see eq.~(3.39) in  Ref.~\cite{Jadach:2015mza},
reads as follows:
\begin{equation}
\begin{split}
&\sigma_{\nlo}^{\msbar}= 
\int d x_\F d x_\B dz\; dx\; \delta_{x=z x_\F x_\B}
\Big\{
 \delta_{1=z} W^{(1)}_{NLO}|_{k_1=0}\;
 d\sigma_0(szx,\hat\theta)\; J_{LO}
\\&
+ \Big(  W^{(1)}_\nlo\; (sx,\alpha,\beta, \Omega)\; J_{NLO}
       - \; J_{LO}
  \Big)
  \big( d^3\rho_1^\F +d^3\rho_1^\B \big)
  \delta_{1-z=\alpha+\beta}
\Big\}
\\&\times
f^{\krk}_q(sx, x_{\F}) f^{\krk}_{\qb}(sx, x_{\B}).
\end{split}
\label{eq:MC_Master}
\end{equation}

The matching between eq.~(\ref{eq:MsbarMaster}) and eq.~(\ref{eq:MC_Master})
results in fixing the form of the MC correcting weight:
\begin{equation}
   W^{(1)}_{NLO}(k_1) = (1+\Delta_{VS})\; 
   \frac{d^5\sigma_1^\nlo\; (sx,\alpha,\beta, \Omega)}%
        {d^5\sigma_1^\F +d^5\sigma_1^\B}.
\label{eq:TWnlo}
\end{equation}
The same matching also provides
the unambiguous relation between PDFs in the $\msbar$ and $\krk$.
In the $\krk$ scheme the entire $\sim \delta(k_{1T}^2) \Sigma_q(z)$ 
is eliminated (modulo ${\cal O}(\alpha_s^2) $ terms)
thanks to the assignment $\hat{s} \equiv s x_{F} x_{B} =\mu^2$
and redefinition of the PDFs
\begin{equation}
\label{eq:MSnar2MC}
 f^{\krk}_{q,\qbar}(   \mu^2,x) =  
 \int dz dx' \delta(x-zx')
 [ \delta(1-z) + \Sigma_q(z) ]_{\hat{s}=\mu^2}\;
 f^{\msbar}_{q,\qbar}(   \mu^2,x'),
\end{equation}

A few remarks are in order:
The term similar to the $\Sigma_{q}(z)$ function is completely absent
in the distribution (\ref{eq:MC_Master}) 
for any kind of parton shower with the NLO corrected hard process.
In the \krknlo\ method it is absorbed in the redefined PDF.
In other matching schemes like \mcatnlo~\cite{Frixione:2002ik}
and \powheg~\cite{Frixione:2007vw} 
this term is incorporated into PDFs by the ``in flight'' transformation
done on the PDFs {\em inside} the MC program during the event generation.
In the \krknlo\ method the same transformation is performed on PDFs
outside the MC program.
Consequently, the process-independence of the $\Sigma_{q}(z)$ function
is very important for the \krknlo\ method and not so important for
the other matching methods%
\footnote{%
However, keeping this transformation {\em outside} the MC makes sense, 
because ``in flight'' transformation of PDFs complicates significantly MC
program and also might be the source of the annoying negative MC weights.
}.
In the above it was assumed that LO MC was identical with
the sum of two CS dipoles. 
In a more general case the denominator of eq.(\ref{eq:TWnlo})
is $d^5\sigma^{\lo}_{q\bar{q}}$ generated in the PS MC
(not necessarily equal to sum of two CS dipoles).
However, the finitness of $W^{(1)}_{NLO}(k_1)$ requires that 
this $d^5\sigma^{\lo}_{q\bar{q}}$
has exactly the same soft and collinear limits as the sum of two CS dipoles.

Having shown the critical role of the $\Sigma_{q}(z)$ function
in the $\krknlo$ matching scheme,
before analysing its process independence (universality), 
let us look more precisely where from it came in our particular DY case.
It is born out from partial integration over the distribution of
the sum of two CS dipoles in $d=4+\veps$ dimensions:
\begin{equation}
 \rho^{CS}_{q\bar{q}\to V}(k_1,\epsilon)=
  \frac{ 2C_F \alpha_s}{\pi}
  \frac{ (4\pi)^{-\epsilon}}{\Gamma(1+\epsilon)} 
  \bigg( \frac{s_1 \alpha\beta}{\mu^2} \bigg)^{\epsilon}
  \frac{1+z^2+\veps(1-z)^2}{2\alpha\beta}\;
  \frac{d\sigma_0}{d\Omega}(zs_1,\theta),
\end{equation}
where $z=1-\alpha-\beta$. 
In the CS subtraction scheme this distribution
{\em is added} in the integrated form 
in $d=4+\veps$ dimensions to NLO virtual corrections 
and {\em subtracted} in $d=4$ dimensions from the real NLO distributions.
As it is well known in the NLO real+virtual distribution 
in the dimensional regularization
remains uncanceled single pole term times LO kernel, 
which in our particular case is
\begin{equation}
2\Lambda^{\msbar}_{q\fr q}(\veps,z)
 =\frac{\alpha_s}{\pi}
  \frac{(4\pi)^{-\veps}}{\Gamma(1+\veps)}
  \frac{1}{\veps} 
  C_F \frac{1+z^2+\veps(1-z)^2}{1-z}.
\end{equation}
In the $\msbar$ scheme this kind of terms, soft collinear counterterms (SCTs),
are simply subtracted%
\footnote{And are replaced by the PDFs in the $\msbar$ scheme.}.
It makes sense to combine CS dipoles with SCTs into a single object,
which is upon (partial) phase space integration in $d=4+\veps$ dimensions
combined with standard virtual corrections.
In our case the above combination is:
\begin{equation}
\begin{split}
\Reu_q(z,\veps) 
&= \int d\alpha d\beta\; d\Omega\;\delta_{1-z-\alpha-\beta}\; 
\rho^{CS}_{q\bar{q}\to V}(k_1,\epsilon)
- 2\Lambda^{\msbar}_{q\fr q}(\veps,z)
\\&
=S_q(\veps) \delta(1-z) +\Sigma_q(z).
\end{split}
\label{eq:SandSigma}
\end{equation}
The above explains clearly the origin of the  $\Sigma_{q}(z)$ function
in the final NLO result in the $\msbar$ scheme
and its relation to the CS dipoles.
The split between two parts of $\Reu_q(z,\veps)$ is unambiguous
due to the requirement that $\Sigma_q$-like part obeys momentum sum rule --
so in fact there is a one to one correspondence 
between $S_q(\veps)$ and $\Sigma_{q}(z)$ functions and CS dipoles.
N.B. The cancellation of $\veps$ poles occurs entirely in one place,
that is between $S_q(\veps)$
and virtual loop corrections from Feynman diagrams.

Let us stress again that the minimal requirements of the \krknlo\ scheme
to work is that single real parton emission distribution
in $d=4$ dimensions for the sum of CS dipoles on one hand
and for the same distribution of any modern LO PS on another hand,
has the same correct soft collinear limit.
In view of that,  in our quest for process independence of the ${\bf K}$-matrix,
{\em we are going to focus on the freedom in the choice of CS dipoles},
because it translates into the shape of the $\Sigma$-like functions
and $\Kbbm$-matrix elements.

Generalising eq.~(\ref{eq:SandSigma}) to an arbitrary process,
for each NLO splitting $K\fr I,\;  K, I = q, \bar{q}, G$ in the NLO process
the following component is present in the final CS NLO distributions:
\begin{equation}
\begin{split}
&\Reu_{K\fr I}(z,\veps) 
= \int d\alpha d\beta\; d\Omega\;\delta_{1-z-\alpha-\beta}\; 
\sum_S \rho^{S}_{K\fr I}(k_1,\epsilon)
    - \Lambda^{\msbar}_{K\fr I}(\veps,z)
\\&~~~~~~~~~~~~~~~~
=S_{K\fr I}(\veps) \delta(1-z) +\Sigma_{K\fr I}(z,\mu_F),
\\&
\Lambda^{\msbar}_{K\fr I}(\veps,z)
 =\frac{\alpha_s}{\pi}
  \frac{(4\pi)^{-\veps}}{\Gamma(1+\veps)}
  \frac{1}{\veps} 
  P_{K\fr I}(z,\veps),
\end{split}
\label{eq:defSigma}
\end{equation}
where $I$ is the emitter, $K$ results from the splitting and $S$ is the spectator%
\footnote{The $S$-dependent colour factor is temporarily omitted. We shall show
   that it cancels out due to colour conservation and spectator independence
   of the modified dipoles.}.

Our reasoning will be now the following:
\begin{itemize}
\item
First of all, the case when both $I$ and $K$ are in the final state (\FF)
is for us uninterestingly trivial.
The integration over dipole for fixed $z\neq 0$ gives $\Sigma_{K\fr I}(z)=0$.
$S_{K\fr I}(\veps)$ gets combined with virtual corrections, 
such that CS dipoles do not need any modification.
\item
Then, the most important modification of the CS scheme is needed 
in case of the final state emitter $I$
and initial state spectator $K$  (\FI)%
\footnote{This case is already present in the DIS process.}.
In the original CS scheme $\Sigma_{K\fr I}(z)$ gets convoluted with PDFs 
and the LO process and the $z$ integration cannot be separated.
Clever modification of the kinematic
mappings in these dipoles will make
the $z$ integration to decouple from PDFs and the LO process, as in the \FF\ case.
\item
Next, we are left only with dipoles with the emitter $I$ in the initial state
and spectator $S$ either in the initial or final state (\II\ or \IF).
We will modify CS dipoles such that $\Sigma_{K\fr I}(z)$ is exactly the same
in both cases.
\item
Finally, $\Sigma_{K\fr I}(z)$ depends also on the combination
of $\ln(2p_I\cdot p_S/\mu_F^2)$ with nontrivial colour coefficients.
We are going to show how to choose $\mu_F^2=\hat\mu_F^2$
in order eliminate this component for an arbitrary process.
\end{itemize}
Once all the above is done, the transformation matrix for PDFs
from $\msbar$ to $\krk$ scheme is given by
\begin{equation}
  \Kbbm_{K\fr I}(z) = \Sigma_{K\fr I}(z,\mu_F)|_{\mu_F^2=\hat\mu_F^2}
\end{equation}
and is process independent.

Finally, let us
remind the reader that the physical meaning of $\Sigma_{q}(z)$ 
is known since pioneering works of Alterelli et.al.~\cite{Altarelli:1979ub}
where it was traced back to the difference between
the upper phase space limit (factorization scale)
being the maximum transverse momentum in PDFs of the $\msbar$ 
and the total available energy in the real world of the hard process.
Obviously, the PDFs of the $\krk$ scheme represent the second, physical, case.

\begin{figure}
\centering
{\includegraphics[width=90mm]{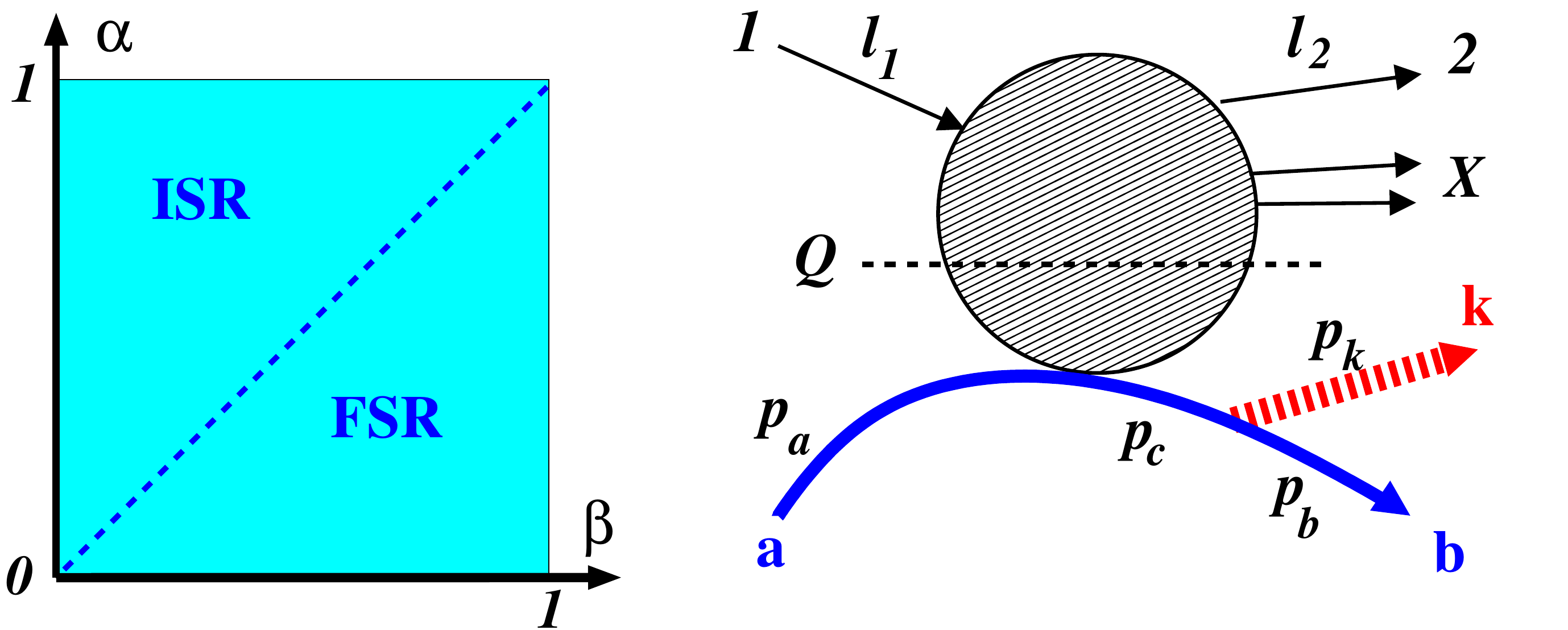}}
\caption{\sf Kinematics and Sudakov plane for \FI\ dipole.}
\label{fig:DISsudak}
\end{figure}

\section{Dipoles with final state emitter and initial state spectator}
It is natural to expect that in the \FI-type dipoles,
with the final state emitter and initial stated spectator,
the integration over dipole internal (Sudakov) variables decouples from 
the factorised LO differential cross section and PDFs,
as it is the case of \FF-type dipoles 
with both emitter and spectator in the final state.
However, it is not the case for the \FI-dipoles 
in the CS work~\cite{Catani:1996vz}.
This is the most sticky issue preventing universality of the $\Kbbm$ transformation,
hence in the following we are going to indicate how to solve this problem,
while fine details will be presented in Ref.~\cite{jadach:2020-next}.

Fig.~\ref{fig:DISsudak} illustrates the kinematics of the \FI\ dipole.
Sudakov variables for the dipole phase space are introduced as follows%
\footnote{This is parametrisation of the CS work~\cite{Catani:1996vz}.
  However it was know earlier, see Ref.~\cite{bhlumi2:1992}.
}:
\begin{equation}
\begin{split}
&p_k=\alfb\; p_a +\betb\; p_b +p_k^T,\quad
\alfb  = \frac{p_k \cdot p_b}{p_a \cdot p_b},\;
\betb  = \frac{p_k \cdot p_a}{p_a \cdot p_b},
\\&
\alpha= \frac{\alfb}{1+\betb},\;
\beta = \frac{\betb}{1+\betb},\;\;
\max(\alpha,\beta) \leq 1,\;\;
\\&
Q=p_b+p_k-p_a,\quad |Q^2|=2p_ap_b\; \frac{1-\alpha}{1-\beta},
\end{split}
\label{eq:DISsudakov}
\end{equation}
The corresponding differential cross section with clear factorization
into the LO process and the dipole radiation parts reads%
\footnote{Colour correlation factor is omitted for the sake of simplicity.}:
\begin{equation}
\begin{split}
& d\sigma^{a}_{bk}
=d\Phi_{4+2\veps}(p_k)\;
 \frac{1}{2 p_b p_k}
  8\pi \mu^{-2\veps} \alpha_s P^*_{b\fr c}(\alpha,\beta)\;
 \frac{p_a \tilde{p}_b}{p_a (\tilde{p}_b-p_k )}
\\&~~~~~~~~~~~\times
\bigg\{
 \frac{1}{s} 
 d\Phi(l_1+ \tilde{p}_a ; \tilde{p}_b, l_2,X) \;
 |\Meu(l_1, \tilde{p}_a ; \tilde{p}_b, l_2,X) |^2
\bigg\}_{d=4+2\veps}
\\&
=\frac{\alpha_s}{2\pi }\;
 \Big(\frac{Q^2}{4\pi \mu^2}\Big)^\veps
 \frac{1}{\Gamma(1+\veps)} 
 \frac{d\Omega^{n-3}(p_k^T)}{\Omega^{n-3}}\;
 H_{bc}(\alpha,\beta,\veps)
 \Big\{ d\sigma^{LO}(l_1,\tilde{p}_a;\tilde{p}_b,l_2,X) \Big\},
\\&
H_{bc}(\alpha,\beta,\veps)=
 \Big( \frac{\alpha \beta(1-\beta) }{(1-\alpha)} \Big)^\veps \;
 \frac{P^*_{b\fr c}(\alpha,\beta,\veps)}{\alpha}.
\end{split}
\label{eq:FIdipole1}
\end{equation}
The above distribution is defined in the entire NLO phase space
$p_a+l_1 \to p_b+p_k+l_2+X$.
However,  in the LO part $\{...\}$ the momentum $p_k$ is eliminated
and effective momenta 
$ \tilde{p}_a = (1-\alpha) p_a,\;\; \tilde{p}_b= Q-\tilde{p}_a,\;\; 
\tilde{p}_a^2= \tilde{p}_b^2=0$  are used.
We denote the 1-particle Lorentz invariant phase space integration element
as $d\Phi_{4+2\veps}(p)$
and $d\Phi(l_1+ \tilde{p}_a ;\tilde{p}_b,l_2,X)$  is the multi-particle phase space element.
$P^*_{b\fr c}(\alpha,\beta,\veps)$ is an extrapolation 
of the spin factor of the splitting kernel over the entire Sudakov phase space, 
which has to coincide with the standard splitting kernel in the collinear limit.
It will be defined in the next section.
In the diagonal case $b=c$ it must exclude 
the initial state $1/\beta$ singularity.
Otherwise it can be freely adjusted to our needs.

The above formula clearly illustrates the problem with the \FI\ dipole,
namely the effective centre of the mass energy in the LO part
$s'=2l_1\cdot \tilde{p}_a= (1-\alpha) s$ depends 
on the Bjorken variable $z_B=1-\alpha$.
(It will also enter into the $x$ argument of the PDF.)

Our alternative solution is that instead of the keeping $z_B$ factor 
in the effective beam momentum $\tilde{p}_a$ of the LO part,
it is just ``boosted out''.
Let us explain how it works.
A boost has a nice property of the Jacobian being equal one.
One may also profit from Lorentz invariance of the LO matrix element.
In Fig.~\ref{fig:DISsudak} particles are divided
into two groups, the dipole part $(a,b,k)$ and the LO rest $(l_1,l_2,X)$.
Two groups are connected by the spacelike exchange 
4-momentum $Q=b+k-a = l_l-l_2-X$.
There is an entire family of the reference frames, in which
$Q=(0,0,|Q^2|^{1/2},|Q^2|^{1/2})$ is pointing along $z$-axis
and has zero energy component.
All these frames are connected with boosts in the $x-y$ plane
perpendicular to $Q$.
Such a frame becomes uniquely defined (modulo azimuthal rotation)
using an additional lightlike momentum, and requiring
that it is along the $z$-axis.
Two such frames are important, $QMS_a$ with $p_a$ along $z$-axis
and $QMS_1$ with $l_1$ along minus $z$-axis.

Now, in the frame $QMS_a$, using the $(a,b,k)$ subset we construct 
the $\tilde{a},\tilde{b}$ effective spectator and emitter.
Then, we go to the $QMS_1$ frame (with $l_1$ along $z$-axis)
and perform the active boost $\Lambda$
in the $x-y$ plane perpendicular to $Q$ on the momenta
of the  $\tilde{a},\tilde{b}$, such that%
\footnote{Using a toy Monte Carlo exercise it was checked that
 such a boost always exists.}
\[
  2 l_1 \cdot \Lambda\tilde{p}_a =s.
\]
The momenta of the $(b,l_k,X)$ are unchanged.
Conservation of the 4-momenta
\[
 Q =\tilde{p}_b -\tilde{p}_a =\Lambda\tilde{p}_b -\Lambda\tilde{p}_a 
   = l_l-l_2-X
\]
holds, because the $\Lambda$ transformation does not change $Q$.
The resulting momenta $\Lambda\tilde{p}_a,\Lambda\tilde{p}_b, l_l,l_2,X$ 
are now ready to be plugged into the LO matrix element. 
(Of course, one may finally transform them to the CMS.)
The explicit dependence on $\alpha$ in the LO part 
of the factorization formula is removed!
In the phase space integration of eq.~(\ref{eq:FIdipole1})
we introduce change of the variables
\[
  l_1=\Lambda l'_1,\quad l_2=\Lambda l'_2,\quad X= \Lambda X'.
\]
and using phase space invariance under Lorentz transformation
eq.~(\ref{eq:FIdipole1}) turns into
\begin{equation}
\begin{split}
& d\sigma^{a}_{bk}
=\frac{\alpha_s}{2\pi }\;
 \Big(\frac{Q^2}{4\pi \mu^2}\Big)^\veps
 \frac{1}{\Gamma(1+\veps)} 
 \frac{d\Omega^{n-3}(p_k^T)}{\Omega^{n-3}}\;
 H_{bc}(\alpha,\beta,\veps)
 \Big\{ d\sigma^{LO}(l'_1,\tilde{p}_a;\tilde{p}_b,l'_2,X') \Big\},
\end{split}
\label{eq:FIdipole1}
\end{equation}
where the condition $ 2 l'_1 \cdot \tilde{p}_a =s= 2l_1 \cdot p_a$ holds,
hence the dipole part decouples from the LO differential cross section 
and can be integrated over analytically, the same way as for \FF\ dipole.
Our goal is achieved.

The following remarks are in order:
We were elaborating on the \FI\ dipole distribution, 
which is added and subtracted in the NLO calculation,
hence it does not change the NLO results.
It is arbitrary to a certain degree and this freedom we have exploited.
In the complete NLO differential cross section the effective rescaling
of the beam energy by the $z_B$ factor is always present.
What we have achieved is that this rescaling is entirely 
encapsulated in the \IF\ dipole
and completely absent in the \FI\ dipole.

\section{Initial state emitter and final state spectator}
The kinematics of the dipole with the initial state emitter 
and final state spectator \IF\ is the same as in Fig.~\ref{fig:DISsudak}
and eq.~(\ref{eq:DISsudakov}) except that the splitting $a\to ck$
is now on the initial leg.
Let us consider separately 
the diagonal splittings $a=b$ with gluon emission
and nondiagonal splitting $a\neq b$, with the quark-gluon transition.

\subsection{Diagonal splittings}
The cases of diagonal splittings $a=b,\; a=q,G$ are special,
because of the presence of the soft singularity in the form of the standard eikonal factor%
\footnote{Omitting for simplicity colour structure.}
$
   \frac{p_ap_b}{(p_kp_a)(p_kp_b)} \sim \frac{1}{\alpha\beta}.
$
In the CS technique such a singularity is split into two parts using
``soft partition functions'' (SPFs) $m_++m_-=1,\; m_\pm\geq 0$:
\[
   \frac{1}{\alpha\beta} =  \frac{1}{\alpha+\beta} \frac{1}{\beta}
                           +\frac{1}{ \alpha+\beta} \frac{1}{\alpha}
   =m_+(\alpha,\beta) \frac{1}{\alpha\beta} +m_-(\alpha,\beta) \frac{1}{\alpha\beta}.
\]
The $m_+/(\alpha\beta)$ part of the eikonal factor is incorporated into 
the \IF\ dipole and $m_-/(\alpha\beta)$ part into the \FI\ dipole.
SPFs are not unique and we are going to examine three choices%
\footnote{Here we always use $m_-=1-m_+$.}:
\begin{equation}
\begin{split}
& m^{(a)}_+(\alpha,\beta)= \theta_{\beta<\alpha},\;
m^{(b)}_+(\alpha,\beta)= \frac{\alpha}{\alpha+\beta},\;
m^{(c)}_+(\alpha,\beta)= \frac{\alpha-\alpha\beta}{\alpha+\beta-\alpha\beta}.
\end{split}
\end{equation}
The important point is that, because  the \FI\ dipole
(thanks to kinematic mapping of the previous section)
does not contribute to the $\Sigma$-function,
by means of manipulating SPFs we may adjust
the $\Sigma$-function from the diagonal \IF\ dipole to be the same
as from the \II\ dipole (our ultimate goal!).

Since the \FI\ and \IF\ dipoles are strongly entangled through the $m_\pm$-functions,
let us write common expression for both of them,
similar to that of eq.~(\ref{eq:FIdipole1}):
\begin{equation}
\begin{split}
& d\sigma^{b \pm}_{ak}
=\frac{\alpha_s}{2\pi }\;
 \Big(\frac{Q^2}{4\pi \mu^2}\Big)^\veps
 \frac{1}{\Gamma(1+\veps)} 
 \frac{d\Omega^{n-3}(p_k^T)}{\Omega^{n-3}}\;
 H^\pm_{aa}(\alpha,\beta,\veps)
 \Big\{ d\sigma^{LO}(l_1,\tilde{p}_a;\tilde{p}_b,l_2,X) \Big\},
\\&
H^\pm_{aa}(\alpha,\beta,\veps)=
 \Big( \frac{\alpha \beta(1-\beta) }{(1-\alpha)} \Big)^\veps \;
 \frac{ m_\pm(\alpha,\beta) \bar{P}_{a\fr a}(z(\alpha,\beta),\veps)}{\alpha\beta},
\end{split}
\label{eq:FIdipole2}
\end{equation}
where the spin numerators of the unregularised diagonal kernels are
\begin{equation}
\begin{split}
&
 \bar{P}_{qq}(z,\veps) = (1-z) \hat{P}_{qq}(z,\veps)
= C_F [ 1+z^2 +\veps (1-z)^2 ],
\\&
 \bar{P}_{GG}(z,\veps) = (1-z)\hat{P}_{GG}(z,\veps)
 = 2C_A \Big(\frac{1}{z} -2(1-z)+z(1-z)^2 \Big)
\end{split}
\end{equation}
and $z(\alpha,\beta)$ must obey the correct collinear limits:
$z(\alpha,0)=1-\alpha$ and $z(0,\beta) = 1-\beta$.
In the present works (in the past as well) we consider three choices :
\begin{equation}
z_A(\alpha,\beta) = 1-\max(\alpha,\beta),\;
z_B(\alpha,\beta) = 1-\alpha,\;
z_C(\alpha,\beta) = (1-\alpha)(1-\beta).
\end{equation}
The upper kinematic limit of the dipole phase space
$\max(\alpha,\beta)\leq 1 $ is always compatible with $z\leq 1$.

In Eq.~(\ref{eq:FIdipole2}) it is always assumed that
in the \FI\ case the mapping $l_1\to l_1',\; l_2\to l_2',\; X\to X'$ 
in order to get $l_1'\cdot \tilde{p}_a =s $ is still to be done,
while for the \IF\ case it is ``ready to go'' with $l_1\cdot \tilde{p}_a = z_Bs $.
However, if we choose $z_A$ or $z_C$ it is then understood that also for the \IF\ case
a similar mapping is done to achieve%
\footnote{This makes easy the integration over the dipole phase space.}
$l'_1\cdot \tilde{p}_a = z_As $ or $l'_1\cdot \tilde{p}_a = z_Cs $.

We have investigated all nine choices of $m_\pm$ and $ z(\alpha,\beta)$
and good choices (compatible with \II) were found 
to be $Aa$, $Ac$, $Ca$ and $Cc$%
\footnote{Details of the calculations will be reported 
  elsewhere~\cite{jadach:2020-next}.},
hence we conclude that for diagonal splitting it is rather easy to achieve
that \IF\ dipoles and \II\ dipoles contribute the same to
$\Sigma_{I\fr I}(z,\mu_F)$ and $\Kbbm_{I\fr I}(z)$.
On the other hand, the singular term $S(\veps)$ in eq.~(\ref{eq:defSigma}),
to be combined virtual corrections, may vary freely with the type of
the dipole.

\subsection{Non-diagonal \IF\ dipoles -- the problem and workaround }
In the \IF\ CS dipoles for non-diagonal splittings $a\neq b,\; a=q,G$
(quark-gluon transitions) the soft singularity is absent 
-- only the collinear singularity is present -- 
the use SPFs is in principle not needed.

Unfortunately, from the straightforward analytical calculations
we get slightly different
$\Sigma_{K\fr I}(z,\mu_F)|_{z \neq 1},\; K \neq I$ for \IF\ dipoles
than for \II\ dipoles for all choices of $z=z(\alpha,\beta)$
defined in the previous subsection.
The difference can be traced back to the upper phase space limit:
$\max(\alpha,\beta)\le 1$ versus $\alpha+\beta\leq 1$%
\footnote{One may map $(\alpha,\beta)\to (\alpha',\beta')$
  such that $\alpha'+\beta'\leq 1$, however, the Jacobian in $d$-dimension
  will cause that the problem is back.}.

The simplest workaround is to split \IF\ non-diagonal dipoles 
into two parts using again SPFs as in the diagonal cases:
\[
 H^{\pm}_{c\fr a}(\alpha,\beta,\veps) 
   = m_\pm(\alpha,\beta) \frac{1}{\beta}
     P_{ca}(z,\veps)\big|_{z=z(\alpha,\beta)},\quad
 c\neq a,
\]
and treat $H^{-}_{c\fr a}$ as additional 
(non-singular) dipoles in the \FI\ class,
decoupled from the LO part and PDFs and
not contributing to $\Sigma_{K\fr I}$.

We have checked that using the above workaround,
the compatibility of \IF\ and \II\ dipoles
is obtained for $q\fr G$ and for $G\fr q$ dipoles
for  $m_\pm^{(a)}$ and $z_A$.
Moreover, the same positive conclusion was obtained
for the combined use of $z_C$ 
and yet another SPF $m_+^{(d)}=1- \beta$.

Altogether, we find that at the expense of introducing
additional non-singular \FI\ dipoles, one can obtain
equality of $\Sigma_{K\fr I}(z,\mu_F)|_{z \neq 1}$ also for
non-diagonal splittings  $K \neq I$.

In this way, we have shown that thanks to judicious choice of
the dipole distributions we are much closer
to the claim the $\Kbbm_{K\fr I}(z)$
matrix is the same independent of whether 
it was obtained from \II\ or \IF\ dipole.

\section{Zeroing the collinear remnant ${\bf P}$}
The role of the term%
\[
   P_{ij}(z) \ln\frac{\hat{s}}{\mu_F^2}
\]
present in the $\Sigma$-function%
\footnote{Sandwiched between the PDF and the LO cross section.}
of eq.~(\ref{eq:CSmaster}) of our introductory DY example
is to keep the factorization scale in PDF to be equal $\hat{s}$.
Any variation of $\mu_F$ in PDFs is compensated by this term,
such that overall dependence on $\mu_F$ in NLO expression cancels
up to \order{\alpha^2}.
It is therefore logical and convenient to set $\mu_F=\hat{s}$
both in the PDF and in the above term, eliminating it completely.
The absence of the above term is also mandatory for the \krknlo\ method
with a single multiplicative MC weight to work.

The above method of eliminating the troublemaking term works well
in DY or DIS process with only two coloured legs.
In the general case the ${\bf P}$-matrix collinear remnant term in the NLO
final result of the CS method reads:
\begin{equation}
\begin{split}
&\sigma_{ab}^{col.rem.} =
\int dx_adx_b\;   
f_b(\mu_F,x_b)\; f_a(\mu_F,x_a)\;
\Big\{
     d\sigma^{Born}_{a,b}(p_a,p_b) + 
\\&
+\sum_{a'} \int dx\;
\Big\langle\;
\frac{\alpha_S}{2\pi} P_{aa'}(x) \Big[ 
\sum_i \frac{T_i\cdot T_{a'}}{T^2_{a'}}\ln\frac{\mu_F^2}{2xs_{ai}}
     + \frac{T_b\cdot T_{a'}}{T^2_{a'}}\ln\frac{\mu_F^2}{2xs_{ab}}
\Big]
\\&~~~~~~~~~\times
d\sigma^{Born}_{a',b}(xp_a,p_b) \; 
\Big\rangle_{color}
+\dots
\Big\},
\label{eq:Pinit}
\end{split}
\end{equation}
where  the summation over $i$ and $b$ is the summation over spectators
and it collects all such logs of many variables $s_{ab}=2p_ap_b$.
Obviously, it is not possible to kill all of them at once
by equating $\mu_F^2$ to one of them.

However, there is a possibility of finding out at each point
of LO phase space (with all $s_{ab}$ defined)
a unique value of $\hat\mu_F$ which renders 
the above entire ${\bf P}$-matrix equal zero. 
Let us show to achieve that.

Using colour conservation 
$ \langle\; T_{a'}+T_b+\sum_i T_i\; \rangle_{color} =0$
and evolution equations for $f_a(\mu,x)$
we obtain easily the following identity:
\begin{equation}
\begin{split}
&\sigma_{ab}^{col.rem.} =
\int dx_adx_b\;   
f_b(\mu_F,x_b)\; f_a(\hat\mu_F,x_a)\;
\Big\{
     d\sigma^{Born}_{a,b}(p_a,p_b)+
\\&
+\sum_{a'} \int dx\; \frac{\alpha_S}{2\pi} P_{aa'}(x) 
\Big\langle\;
 \Big[ 
\sum_i \frac{T_i\cdot T_{a'}}{T^2_{a'}}\ln\frac{\mu_F^2}{2xs_{ai}}
     + \frac{T_b\cdot T_{a'}}{T^2_{a'}}\ln\frac{\mu_F^2}{2xs_{ab}}
+ \ln\frac{\hat\mu_F^2}{\mu_F^2}
\Big]
\\& ~~~~~~~~~~\times
d\sigma^{Born}_{a',b}(x x_a p_1,x_b p_2) \; 
\Big\rangle_{color}
+\dots
\Big\}
\end{split}
\end{equation}
Since $\mu_F^2$ is a local dummy parameter in the above expression
(colour conservation!), we may substitute $\mu_F^2=2xs_{ab}$,
and solve for $\hat\mu_F$ the following equation:
\begin{equation}
\begin{split}
&\sum_{a'}\! \int_0^1\!\!\!\! dz P_{aa'}(z) 
\sum_i 
\ln\frac{s_{ab}}{s_{ai}}
\Big\langle
\frac{T_i\cdot T_{a'}}{T^2_{a'}}
d\sigma^{Born}_{a',b}(zp_a,p_b)  \Big\rangle_{color} +
\\&~~~~~~~~~~
+\sum_{a'}\! \int_0^1\!\!\!\! dz P_{aa'}(z) 
d\sigma^{Born}_{a',b}(zp_a,p_b)
\ln\frac{\hat\mu_F^2}{2zs_{ab}}
\equiv 0.
\end{split}
\label{eq:Pfinal}
\end{equation}
The effective scale $\hat\mu_F$ to be inserted in the PDF in the \krk\ scheme
can be calculated numerically (1-dim. integral over $z$) 
at each point of the Born phase space, 
$h_1+h_2\to p_a+p_b\to 1+2+\dots m$,
or even analytically in some simpler cases.
Of course, for the other PDF $f_b$ a similar independent equation
has to be solved and the resulting $\hat\mu_F$ will be inserted into $f_b$.

In the construction of all new CS dipoles in the previous sections
we have ignored the role of the colour factors.
They enter for a given $a\to a'$ splitting
within the summation over all spectators
\[
 \sum_{S=i,b} 
 \Big\langle
 \frac{T_S\cdot T_{a'}}{T^2_{a'}}
 \;\dots\;
 \Big\rangle_{color},
\]
in a similar way as eqs.(\ref{eq:Pinit}-\ref{eq:Pfinal}).
Now, thanks to the achieved independence of the partly integrated%
\footnote{The integrated contribution for fixed $z\neq 0$.}
modified dipoles on the type of spectator $S=i,b$
and using colour conservation, we see that
the above colour factor factorizes out and gets reduced to unity.
This is yet another important profit from our modification of the CS dipoles!

Eliminating the collinear remnant, ${\bf P}$, 
in the NLO differential distribution
was the last obstacle on the way to making the $\Kbbm_{K\fr I}(z)$ matrix
process independent (universal).

We did not provide in this paper explicit expressions
for the transition matrix $\Kbbm_{K\fr I}(z)$ 
for transforming PDFs from the $\msbar$ to the \krk\ scheme
because they are the same as in eq.~(4.3) of ref.~\cite{Jadach:2016acv}, 
where they were calculated for the Drell-Yan process
and now are applicable to any process.

\section{Summary}
In our analysis we have exploited the machinery of the Catani-Seymour subtraction
scheme to examine the question of universality of the PDFs
in the \krk\ factorization scheme, originally defined and used for
the Drell-Yan type production of heavy colourless bosons.
The transition matrix $\Kbbm_{K\fr I}(z)$ 
for transforming PDFs from the $\msbar$ to the \krk\ scheme
is closely related to partially integrated CS dipoles,
while the MC weight of the \krknlo\ matching scheme also reflects
the shape and normalization of the CS dipoles.
The original dipoles of the CS work do not lead to universality of $\Kbbm_{K\fr I}(z)$.
However, we have shown that one may modify CS dipoles in such a way
that they provide a process independent $\Kbbm_{K\fr I}(z)$.
The key features of the new CS dipoles are that dipoles with final emitter
and initial spectators decouple kinetically from PDFs 
and LO differential distributions (thanks to a new mapping of the dipole kinematics)
and that the remaining dipoles with an initial emitter yield the same
contribution to $\Kbbm_{K\fr I}(z)$ for spectators in the initial and final state.
Full details of the calculations related to new CS dipoles
will be reported elsewhere~\cite{jadach:2020-next}.

\vspace{10mm}
\noindent
{\bf\large Acknowledgments}

The author is indebted to prof. B.F.L. Ward for reading the manuscript 
and to prof. W P\l{}aczek for the valuable criticism and useful corrections.

\newpage

\end{document}